\documentclass[12pt,preprint]{aastex}
\received{}
\accepted{}
\journalid{}{}
\articleid{}{}

\def\sun{\ifmmode\odot\else$\odot$\fi}

\def\degns{\ifmmode^\circ\else$^\circ$\fi}
\def\deg{\ifmmode^\circ\else$^\circ$\fi\ }

\def\h2co{\hbox{H$_2$CO}}

\def\H76{\hbox{H~76$\alpha$}}
\def\H2O{\hbox{H$_2$O}}
\def\h2o{\hbox{H$_2$O}}
\def\degns{\ifmmode^\circ\else$^\circ$\fi}
\def\deg{\ifmmode^\circ\else$^\circ$\fi\ }

\def\kms{~km~s$^{-1}$}
\def\kmss{~km~s$^{-1}$~}

\def\etal{et~al. }

\def\NH3{\hbox{NH$_3$}}

\def\wte{W\thinspace{28} }
\def\wtens{W\thinspace{28}}

\def\j18{J1801$-$231}
\begin{document}

\title{The Size of the Extragalactic Source \j18 
and the Association of Pulsar PSR B1758$-$23 with the
Supernova Remnant \wtens}

\author{M. J Claussen\altaffilmark{1},\,\,W. M. Goss\altaffilmark{1},\,\, 
K. M. Desai\altaffilmark{2},\,\,and C. L. Brogan\altaffilmark{1}} 
\altaffiltext{1}{National Radio Astronomy Observatory (NRAO)
Array Operations Center, P.O. Box O, Socorro, New Mexico 87801, USA}
\altaffiltext{2}{Renaissance Technologies} 

\slugcomment{Accepted for publication in {\it The Astrophysical Journal}}
\begin{abstract}

We have used the NRAO Very Large Array (VLA) in conjunction with the
Very Long Baseline Array (VLBA) Pie Town antenna as a real-time 
interferometer system to measure the size of the extragalactic source 
\j18 as a function of frequency from 1285 to 4885 MHz.  These observations
were made in an attempt to determine the effect interstellar scattering has
on the observed sizes of OH (1720 MHz) masers in the nearby 
(d $=$ 2.5 kpc) supernova remnant \wtens.  

The observations clearly show that \j18 displays angular broadening due
to turbulence in the Galaxy's interstellar medium.  The minimum distance of 
the nearby (two arcminutes from \j18) pulsar PSR B1758$-$23 is constrained to 
be 9.4$\pm$2.4 kpc.  This value is based on both the measured
size of 220 mas for \j18 at 1715 MHz and the temporal broadening 
of the pulsar.  A single thin scattering screen along the line 
of sight to the \wte OH(1720 MHz) masers must be at 4.7$\pm$1.2 kpc for this minimum
pulsar distance. The screen may be placed closer to the Earth, but for reasonable
values of the pulsar distance (i.e., the pulsar is within the Galaxy), this
choice leads to a negligible scattering contribution to the sizes of the masers.

Thus the OH(1720 MHz) masers, at a distance of 2.5$\pm$0.7 kpc, 
are unaffected by interstellar scattering, and the measured maser sizes must be 
intrinsic.  Our measured upper limits to the size of the pulsar itself are 
consistent with the distance estimates to the pulsar and the scattering screen.

\end{abstract}


\section{Introduction}

The \wte supernova remnant (SNR) lies in the direction of
the Galactic Center: $(l,b)=(6.8,-0.06)$.  
The distance to \wte is somewhat uncertain.  Since
the distance to the SNR is important to the conclusions of the current
study, it is useful to examine previous estimates of the distance found
in the literature.
Many authors have used the $\Sigma-D$ relation to derive a distance
to \wte near $\sim$2 kpc (Milne 1970; Clark \& Caswell
1976; Goudis 1976; Milne 1979); however, as discussed by Kaspi \etal
(1993), this method is extremely uncertain.  The furthest distance
estimated in the literature is 3.6 kpc by Lozinskaya (1974), based on
H$\alpha$ measurements and assuming an LSR velocity near +18\kmss
for \wtens.  Velazquez \etal (2002) estimate a distance of 1.9$\pm$0.3 kpc
adopting +7\kmss for the LSR velocity, and assuming a standard circular
rotation model (this is the near kinematic distance; 15 kpc is the
far kinematic distance). Frail, Kulkarni, \& Vasisht (1993) adopt a distance of 3 kpc,
based on HI absorption, assuming an LSR velocity of 17.6\kms. While the
weight of the evidence for distance estimates to \wte favors a lower
distance ($\sim$ 2 kpc; the argument of Velazquez \etal 2002 for an LSR 
velocity of +7\kms is especially strong), we adopt a conservative estimate
for the distance to \wte of 2.5$\pm$0.7 kpc.

The 60,000
year old pulsar PSR~B1578$-$23 lies in the same direction as \wte but is
located outside the SNR, approximately three arcminutes to the north 
of its bright radio continuum edge.  A radio continuum source, \j18 
(also known as 1758$-$231), lies within two arcminutes of this pulsar.  
\j18 is assumed to be an extragalactic source, coincidentally along
the line of sight to the pulsar, based on measurements of neutral hydrogen
absorption (Frail, Kulkarni, \& Vasisht 1993).  Using observations of the 
pulsar and the neighboring extragalactic source, Frail et al. argued in favor of the 
association of the pulsar 
and the SNR.  Kaspi et al. (1993) disagreed, suggesting that the pulsar 
was much more distant.  The discussion of Frail \etal was based upon the
similarity of HI absorption profiles toward the pulsar and extragalactic source.
Kaspi et al. base
their conclusion on the high dispersion measure of the pulsar 
(1074 pc cm$^{-3}$), and the argument that no compact HII region or 
cloud appears to lie along the line of sight to the pulsar which could 
account for such a high dispersion measure.

Several OH (1720 MHz) masers associated with the \wte SNR (Frail, Goss, 
\& Slysh 1994) have 
been studied at high angular resolution with the Very Long Baseline
Array (VLBA) and MERLIN by Claussen \etal (1999a).  Though there have been 
only a few OH(1720 MHz) masers in \wte observed at 10 mas resolution, 
those masers are resolved and have sizes ranging from 50 to 100 
milliarcseconds (mas).  Claussen \etal discussed the effects of interstellar 
scattering on the sizes of the masers and the constraints that could be
placed on such scattering sizes based on available information about the
size of \j18, the pulsar, and the pulse broadening of the pulsar.
Interstellar scattering of an extragalactic source results in angular 
broadening while interstellar scattering of a pulsar's radio emission
can also result in pulse broadening, an increase in the apparent width 
of a pulsar's average pulse profile beyond its intrinsic width.  The degree 
of angular broadening for the masers and the extragalactic source and the 
degree of pulse broadening for the pulsar all depend in different ways 
upon the relative geometry of the observer, scattering material, and 
sources.  Measurements of these scattering effects can constrain the 
distribution of the scattering material and can be used to estimate, 
for example, the unscattered sizes of the masers.

In this paper, we present angular broadening measurements of the 
extragalactic source 
\j18 as a function of observing frequency from 1285 MHz to 4885 MHz. 
We used the NRAO Very Large Array (VLA) in its most extended configuration 
({\bf A}) linked, in real time, with the Pie Town (Pt)
antenna of the Very Long Baseline Array (VLBA).
The questions to be addressed by these measurements are (1) the relative distances
of the pulsar and the supernova remnant (which we assume harbors the OH
masers), (2) the effects of scattering on the observed size of the masers and
the pulsar, and (3) the distance to the scattering screen.

\section{Observations and Data Reduction}

\subsection{Observations}
The VLBA Pie Town (Pt) antenna has been linked in real
time via fiber-optic cable with the VLA (Claussen et al. 1999b; Beresford 2000).  
For purposes of scientific
observations, the data from the Pt antenna is sent to the
VLA correlator just like any other VLA antenna.  For sources with 
declinations $>$ 40\deg, the angular resolution
in two dimensions is improved by approximately a factor two compared with using the VLA
in the {\bf A} configuration.  As the declination of sources tends to
the south, the increase in resolution is realized in only one dimension.

We observed the extragalactic source \j18 on January 15, 2001, with the
VLA+Pt link.  We performed five snapshots of the source of approximately
10 minutes each on source at six frequencies: 1285, 1315, 1365, 1665, 1715, and
4885 MHz.  The five snapshots covered an hour angle range of about $\pm$1.5
hours.  At the frequencies 1285 to 1715 MHz, we employed 25 MHz front-end
filters to help avoid radio-frequency interference and minimize the radial
smearing due to the effects of finite bandwidth (see e.g. Bridle \& Schwab 1999,
and references therein).  At 
4885 MHz the filters were set at 50 MHz.  Two independent circular 
polarizations were observed
at each frequency, and subsequently averaged in the imaging process.
The data were edited, calibrated, and imaged in the standard
manner, using the NRAO Astronomical Image Processing System (AIPS).  
Absolute flux calibration was provided assuming that 3C~286 has flux
densities of 15.44, 15.27, 15.01, 13.63, 13.43, and 7.46 Jy beam$^{-1}$ at 
1285, 1315, 1365, 1665, 1715, and 4885 MHz respectively.

Uniform weighting of the {\it u,v} data was used in the imaging to
allow the data from Pt baselines to contribute to the image.  The AIPS task
JMFIT was used to fit two-dimensional gaussians to the brightness distribution
of \j18, and to deconvolve the synthesized beam from those fits.  Table 1 shows a
summary of the observations, imaging, and gaussian fits to the images of
\j18.  Since the sizes of the radio source measured by fitting and deconvolving 
the observed size are nearly circular, the deconvolved sizes reported in Table 1 
are the geometric mean of the two-dimensional gaussian fits.  We also detected 
the pulsar B1758$-$23; the image of the pulsar suffers
from severe bandwidth smearing, since it is two arcminutes from the 
phase center.  Accurate size measurements of the pulsar are thus compromised.
However, we attempted to correct the measured sizes of the pulsar by estimating
the magnitude of the smearing.  At all frequencies, the best 
estimate of the size of the pulsar is $<$ 0.5\ arcseconds.

\section{Results and Discussion}
Figure 1 shows the results of our angular size measurements of 
\j18: a log-log plot of the deconvolved size $\theta$ versus observing 
frequency.  A well-known frequency scaling relation exists for the
measured angular size of a source whose structure is dominated by
interstellar scattering effects (Rickett 1977):
$$\theta_{scattered} \propto \nu^{\gamma}\,\,\,\,,$$
where $\gamma$ takes on a value between $-$2 and $-$2.2 .  
For reference, we have drawn a line with slope $-2$ in Figure 1;
it is clear that the measured sizes shown are consistent with the
effects of interstellar scattering.
  
Under the assumption that the intervening turbulence is confined to
a single thin screen, scattering measurements of galactic and
extragalactic compact sources can completely constrain the location
of the screen.  The assumption that the scattering screen is thin
is equivalent to the statement that the scattering screen has 
a thickness which is much less than the distance to the extragalactic
source (e.g. Trotter, Moran, \& Rodr\'iguez 1998).  Since the extragalactic source is
likely to be at least several megaparsec away, a thickness of several
pc for the screen is consistent with this assumption.  
Temporal broadening of the pulsed emission from a
pulsar and angular broadening of pulsars and masers depend 
differently upon the distribution of scattering material along the
line of sight. Additionally, angular broadening of the extragalactic
source \j18 is sensitive only to the strength of the turbulence,
not to its distribution along the line of sight.

In addition to the angular size measurements of \j18 and the
pulsar PSR B1758$-$23 reported here,
we also use the measured size of the OH masers (Claussen et al. 1999a), and
the measured temporal broadening of the pulse of the pulsar (Kaspi et al. 1993;
Frail, Kulkarni, \& Vasisht 1993)
in order to estimate 1) the distance to the pulsar, 2) the distance to the scattering
screen, and 3) the unscattered size of the OH masers.  In order to make these
estimates, we make the further assumption that the scattering screen is uniform
across both the pulsar and the extragalactic source (the two radio sources are
separated by about 2 arcminutes).  At a distance of 10 kpc, 2 arcminutes corresponds
to about 6 pc.  On pc scales, it is not unreasonable to expect that 
the scattering medium is at least somewhat uniform.  
Lazio et. al (1999) find that the Galactic center scattering region
(covering Sgr A*, OH masers, and extragalactic sources) is inhomogeneous
on scales of order 10 pc, whereas Fey, Spangler, \& Mutel (1989) find that
scattering in the Cygnus region changes by factors of two to five over angular
separations of a few degrees (100---200 pc).
 
If we denote the measured sizes of the extragalactic source, the pulsar,
and the OH(1720 MHz) masers as $\theta_e, \theta_p,$ and $\theta_m,$ 
respectively, the distance to the scattering screen, the masers, and the
pulsar as $d_s$, $d_m$, and $d_p$, respectively, and the temporal broadening 
of the pulsar as $\tau_p$, then three relevant relations can be derived 
(e. g. Britton, Gwinn, \& Ojeda 1998; Frail, Kulkarni, \& Vasisht 1993):
\begin{equation} \theta_e = \theta_p (1 + f_p) \end{equation}
\begin{equation} \theta_e = \theta_m (1 + f_m) \end{equation}
\begin{equation} 1 + f_p = (\theta_e)^2\,\, d_s / (8c\,\, {\rm ln} 2\,\, \tau_p)\,\,\,\, , \end{equation}
where 
$$f_p = { d_s \over  (d_p - d_s)}\,\,\,,\,\,\,\, f_m = {d_s \over  (d_m - d_s)}\,\,\,\, ,$$
and $f_p$ and $f_m$ are dimensionless while $\theta_e, \theta_m$, 
and $\theta_p$ are
in radians, $\tau_p$ is in seconds, and $c$ is the speed of light.

We can rearrange these equations to obtain:

\begin{equation} {\theta_p \over \theta_e} + {d_s \over d_p} = 1 \end{equation}

\begin{equation} {\theta_m \over \theta_e} + {d_s \over d_m} = 1 \end{equation}
\begin{equation} {d_s \over d_p} + { (8c\,\, {\rm ln} 2\,\, \tau_p) \over (\theta_e^2 d_s)} = 1\,\,\,\, . \end{equation}
In distance units of kpc, time units of seconds, and angular units of 
arcseconds, Equation \arabic{equation} becomes:
\begin{equation} { d_s \over d_p} + 2.292\,\, {\tau_p\over (\theta_e^2\,\, d_s)} = 1 \end{equation}
and Equation 4 becomes
\begin{equation} \theta_p = {{ 2.292\,\,\tau_p} \over {\theta_e\,\,d_s}}\,\,\,\, . \end{equation}
Measurements of the three angular sizes $\theta_e, \theta_p$, and
$\theta_m$ and the broadening time $\tau_p$, will thus lead to a determination of
$d_s, d_m$, and $d_p$.

In practice, the angular sizes and the broadening time are measured at 
different frequencies.  Since scattering quantities are frequency dependent, 
we must transform the measurements to a common frequency.  The choice of 
frequency is obviously the frequency of the maser angular 
size measurement, 1720 MHz. The size measurement of the OH masers is 
subject to two different interpretations: scattering due to interstellar
turbulence or large intrinsic maser sizes (Lockett et al. 1999).  We therefore 
scale all measurements to 1720 MHz and apply the above relations at that 
single frequency.

Theoretically, $\tau_p$\ scales with $\nu^{2\,\gamma}$ while
$\theta_e, \theta_m, \theta_p$ scale with $\nu^\gamma$.  As mentioned
earlier, $\gamma$ can take on a value between $-$2.0 and $-$2.2 --- the 
exact value used for $\gamma$\ will have a substantial effect on the derived
quantities.  Uncertainties in the measurement of the sizes and the pulse broadening,
along with the uncertainty in $\gamma$, lead to a range of possible estimates 
of the distances to the pulsar and the scattering screen and the size of
the pulsar.  

Although there is considerable uncertainty as mentioned above, a clear
result can be obtained if we consider the direct measurement of the size
of \j18 at 1715 MHz ($\theta_e = $220 mas).  A plot of the distance to
the pulsar vs the distance to the scattering screen, using this value
for $\theta_e$ in Equation 7 is shown in Figure 2.  The pulse broadening 
time $\tau_p$\ in Equation 7, necessary for the plot, was estimated by 
Frail \etal to be 0.07 seconds at 1670 MHz.  The 1520 MHz pulse profile 
in Kaspi \etal is consistent with a slightly smaller 1670 MHz value in 
the range 0.055 to 0.057 seconds.  Both of these measurements taken 
together would suggest a value for $\tau_p$ at 1720 MHz in the range 
0.048 to 0.051 seconds.  In our analysis we find that there are no
significant differences in our conclusions using this range of values for
$\tau_p$. The plot in Figure 2 uses the mean of pulse 
broadening times, 0.0495 seconds.  Dashed lines in Figure 2 show the same
relation for pulsar distance --- screen distance for $\theta_e = 180$ and
$\theta_e = 260$ mas.  For $\theta_e = 220$ mas, Figure 2 shows a clear 
minimum for the
pulsar distance of about 9.4 kpc, which requires the scattering screen to
be at a distance of 4.7 kpc.  The SNR distance of 2.5 kpc is also plotted on
Figure 2, for reference.

The solid line in Figure 2 is determined directly from Equation 7 with $\tau_p = 
0.0495$ seconds and $\theta_e = 220$ mas, so 
we can analytically determine where the minimum pulsar distance
occurs.  Equation 7 can be written as a quadratic
equation in $d_s$, 
\begin{equation} {d_s}^2 - d_p d_s + \beta d_p = 0\,\,\,\, , \end{equation} where
$$\beta = 2.292 {\tau_p \over {\theta_e}^2}\,\,\,\, .$$
The solutions (for $d_s$) to this quadratic equation are
\begin{equation} d_s = {d_p \pm \sqrt{d_p^2 - 4\beta d_p} \over {2}}\,\,\,\, . \end{equation}
We can determine a minimum value for $d_p$ by taking the derivative ${d d_p} \over {d d_s}$.
This leads to the well-known relation $$d_{p,min} = 2 {d_{s} }\,\,\,\, $$
(e.g. Gwinn, Bartel, \& Cordes 1993).
This minimum value of $d_p$ occurs when $d_p = 4\beta$ and $d_p = 0$. Since
$d_p = 0$ is not an acceptable solution, we find that the minimum
value for $d_p$ is $4 \beta$.  Table 2 lists a few values for $d_{p,min}$ and
the value of $d_s$ where $d_{p,min}$ occurs, using values of $\theta_e$ which bracket 
the errors of our measurements.  The entries in Table 2 show that, even for values of 
$\theta_e$ outside our one-sigma errors, the minimum distance for the pulsar
is well beyond the distance to the \wte SNR.  Based on these values
in Table 2, we can estimate the uncertainty in the 
minimum pulsar distance to be $\pm$2.4 kpc, with a corresponding
uncertainty in the screen distance to be $\pm$1.2 kpc.

With an estimate of the minimum distance to the pulsar and a corresponding
distance to the screen established, we can
use Equation 8 to estimate the scattered size of the pulsar.  This
estimate is 110$\pm$15  mas at 1720 MHz, well below our upper limit for the
size of the pulsar.  Table 2 also lists the values of the pulsar
scattered sizes ($\theta_p$) for different values of $\theta_e$.

It is clear from Figure 2 that smaller values of
the screen distance are possible for a given value of $\theta_e$,
as the pulsar distance increases (to the left of
the minimum pulsar distance in Figure 2).  We can use the solid curve in Figure 2, 
in conjunction with Equation 2, to estimate the size of the OH(1720 MHz) masers
($\theta_m$), based on an extreme value for the pulsar distance.  Selecting
a pulsar distance of 15 kpc, the screen distance would have to be 2.9 kpc,
and thus the ratio of the extragalactic source size to the maser source size
($1 + f_m$) would be $\sim$33, predicting a maser scattering size of $<$ 7 mas.
Placing the screen closer to the Earth along the line of sight (in order to decrease
$f_m$ and increase the scattering size of the masers) requires that the pulsar be 
even more distant, likely outside the Galaxy.

These estimates for the distances of the scattering screen and the pulsar
have two immediate ramifications: (1) the pulsar is {\it not} associated
with the \wte SNR; and (2) the OH(1720 MHz) masers are unscattered, since
placing the screen at 4.7 kpc puts it beyond the distance to the supernova remnant
(2.5 kpc).  Thus the measured angular sizes of the \wte masers as reported by 
Claussen \etal (1999a) (50 --- 100 mas) are intrinsic to the maser rather
than due to interstellar scattering.  Finally, our estimate of the distance
to the pulsar is more consistent with the dispersion measure
of the pulsar as measured by Kaspi et al. (1993) than with a closer
distance ($d >$ 9.5 kpc). 

\section{Conclusions}

We have measured the size of the extragalactic source \j18 at six
frequencies, and find that our size measurements are consistent
with angular broadening due to turbulence in the interstellar
medium.  Together with the size measurements of OH(1720 MHz) masers,
assumed to be associated with the \wte supernova remnant (Claussen
et al. 1999a), and the temporal broadening of the pulsar B1758$-$23
(Frail, Kulkarni, \& Vasisht 1993; Kaspi et al. 1993), the results of the
new size measurements of \j18 imply:

$\bullet$  The minimum distance for the pulsar is 9.4$\pm$2.4  kpc and it is {\it not}
associated with the SNR.  The distance to a single scattering screen
for this minimum pulsar distance is 4.7$\pm$1.2 kpc, and is also beyond the
SNR.

$\bullet$  The scattering screen can be moved placed closer to the Earth by
increasing the pulsar distance.  Even with an extreme distance to the pulsar,
this leads to a negligible scattering contribution to the size of the OH
masers.

$\bullet$  In either of the two possibilities above, the maser sizes must be
little affected by interstellar scattering, and the sizes measured for the masers 
must therefore be intrinsic to the masers.

$\bullet$ When the pulsar is placed at the minimum distance of 9.4 kpc, the 
scattering size of the pulsar, estimated from the \j18 size
measurements, is about 110 mas at 1720 MHz, consistent with our upper 
limit to the pulsar size of 500 mas.

\acknowledgments
We thank Walter Brisken for estimating the size of the pulsar in 
the presence of bandwidth smearing, and we also thank an anonymous
referee for a careful reading of the manuscript and suggestions which
improved the paper.  The National Radio Astronomy 
Observatory is a facility of the National Science Foundation, operated 
under cooperative agreement by Associated Universities, Inc.

\clearpage
\begin{table}
\begin{center} 
\caption{Summary of Observations of \j18}
\begin{tabular}{lccc}
Frequency (MHz) & Beam (mas) & $\theta_{J1801-231}$ (mas) & Total Flux Density (mJy) \\
\tableline
\tableline
1285 & 2660 $\times$ 1290$^1$ & 580$\pm$150 & 54.8 \\
1315 & 2360 $\times$ 680 & 475$\pm$40 & 56.2 \\
1365 & 2140 $\times$ 660 & 500$\pm$40 & 59.6 \\
1665 & 1800 $\times$ 560 & 310$\pm$40 & 51.2 \\
1715 & 1770 $\times$ 540 & 220$\pm$40 & 45.5 \\
4885 &  650 $\times$ 280 & $<$90      & 23.6 \\
\tableline
$^1$ VLA only. 
\end{tabular}
\end{center}
\end{table}
\clearpage
\begin{table}
\begin{center}
\caption{Derived values of minimum pulsar distance, screen distance at which the minimum
pulsar distance occurs, and the pulsar size for this screen distance.}
\begin{tabular} {lccc}
$\theta_e$(mas) & $d_{p,min}$(kpc) & $d_{s}$(kpc) & $\theta_p$(mas) \\
\tableline
\tableline
180 & 14.0 & 7.0 &  90 \\
200 & 11.4 & 5.7 & 100 \\
220 &  9.4 & 4.7 & 110 \\
260 &  6.7 & 3.4 & 130 \\
300 &  5.0 & 2.5 & 150 \\
\tableline
\end{tabular}
\end{center}
\end{table}
\clearpage
\begin{figure}
\plotone{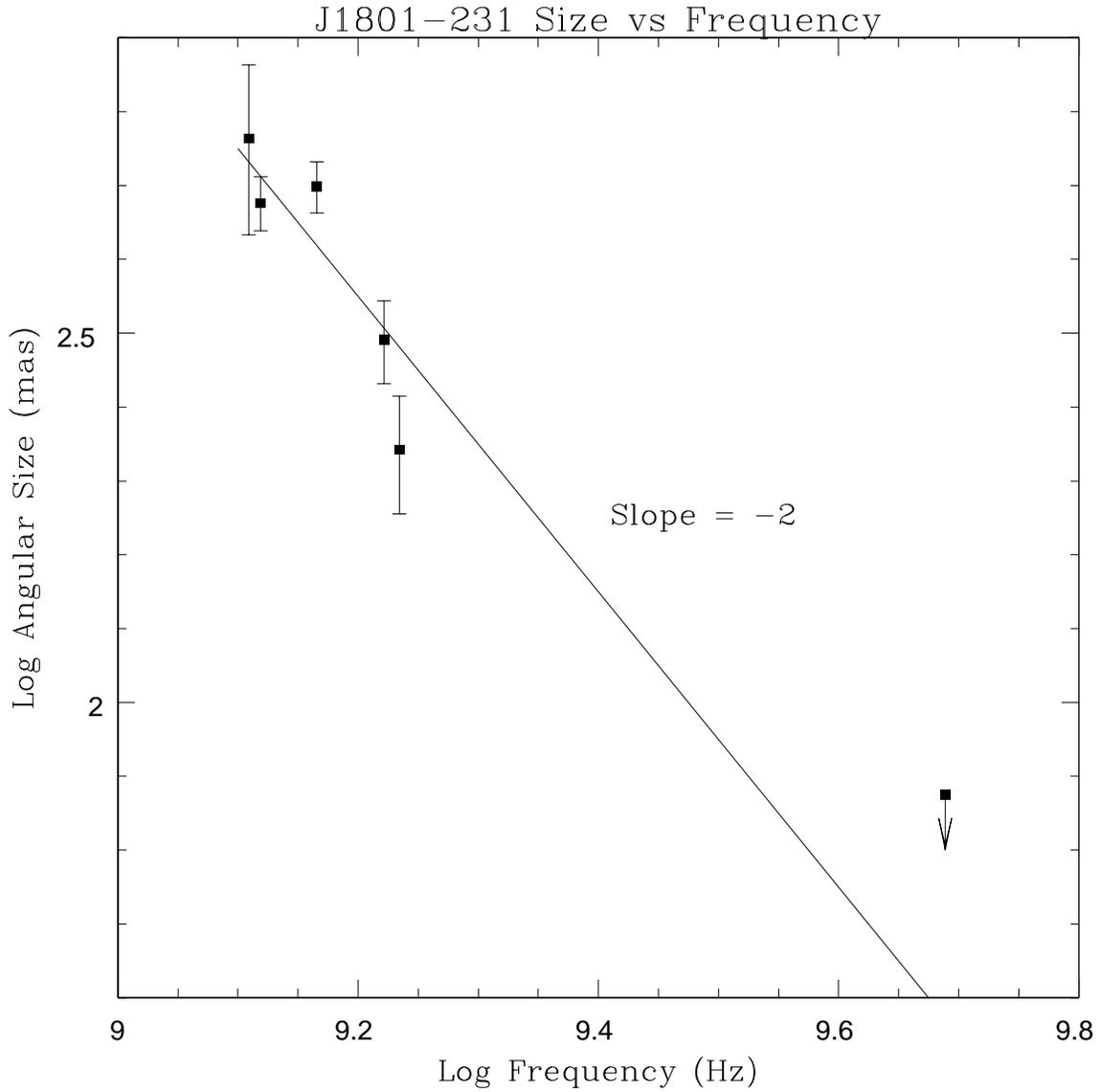}
\caption{A log-log plot of the measured size of the extragalactic
source \j18 vs the observing frequency.  The size of the source
at 4885 MHz (the point at the far right) is an upper limit.  The
solid line is a line with slope $-$2; it is
{\it not} a fit to the data points.}
\end{figure}
\begin{figure}
\plotone{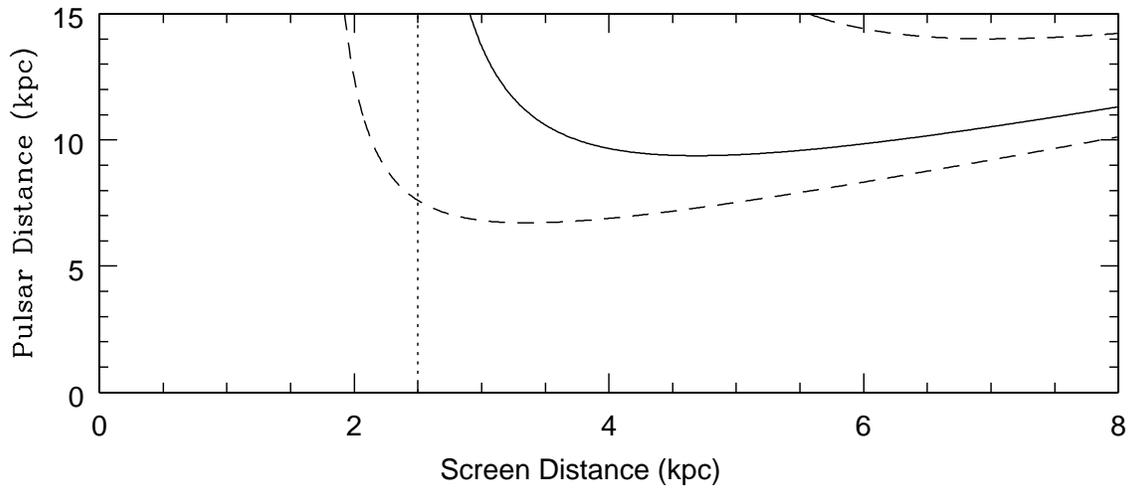}
\caption{Plot of the distance to the pulsar B1758$-$23 vs. 
a single scattering screen's distance based on Equation 7.  The 
solid line assumes a scattering size for \j18 to be 220 mas at 1715 MHz, and
the pulsar temporal broadening is assumed to be 0.0495 seconds. The dashed
lines show pulsar distance - screen distance relations for the scattering
size of \j18 of 180 mas (upper curve) and 260 mas (lower curve).
The dotted line shows the assumed distance to the \wte supernova remnant.}
\end{figure}

\end{document}